# Ab-initio study of quantum oscillation in altermagnetic and nonmagnetic phases of RuO$_2$

Yingliang Huang,[1,2] Junwen Lai,[1,2] Jie Zhan,[1,2] Tianye Yu,[2] Rong Chen,[2,3] Peitao Liu,[1,2] Xing-Qiu Chen,[1,2,*] and Yan Sun[1,2,†]

[1]School of Materials Science and Engineering, University of Science and Technology of China, Shenyang 110016, China

[2]Shenyang National Laboratory for Materials Science, Institute of Metal Research, Chinese Academy of Sciences, Shenyang 110016, China

[3]School of Materials Science and Engineering, Northeastern University, Shenyang 110819, China

E-mail: xingqiu.chen@imr.ac.cn; sunyan@imr.ac.cn

**Abstract**

Altermagnet (AM) is a new proposed magnetic state with collinear antiferromagnetic ground state but presents some transport properties that were only believed to exist in ferromagnets or non-collinear antiferromagnets. To have a comprehensive understanding of the transport properties of AMs, especially from the experimental point of view, a promising altermagnetic metal is crucial. In all the proposed altermagnetic metals, RuO$_2$ has a special position, since it is the first proposed AM with the largest spin splitting and several important altermagnetism featured experiments were first performed based on it. However, a very recent report based on sensitive muon-spin measurements suggest a super small local magnetization from Ru, i.e. a nonmagnetic ground state in RuO$_2$. Therefore, a determination of the existence of the altermagnetic ground state is the basic starting point for all the previously altermagnetic transport properties in RuO$_2$. In this work, we propose to identify its magnetic ground state from the Fermi surface (FS) via the electronic transport property of quantum oscillation (QO). We systematically analyzed the FSs of RuO$_2$ in both nonmagnetic and altermagnetic states via first principles calculations. Our work should be helpful for future experiments on QO measurements to confirm its ground state by the interplay between transport measurements and computations.

## I. BACKGROUND AND INTRODUCTION

In traditional understanding, the collinear magnetic ordered state can be classified into ferromagnet (FM) and antiferromagnet (AFM) according to their ground state and exchange energies. Very recently, the understanding of collinear AFMs was refreshed and a new magnetic state of altermagnet (AM) was proposed from the symmetry points of view[1-7]. Different from traditional AFM with two sublattices connected by time-reversal and translation/inversion[8], the AM follows the rotation symmetry. With the special symmetry, the AM presents as AFM in energy ground state and FM-like spin split in electronic band structure[9-12]. Owing to the spin split, the AM hosts some special transport properties that were only believed to exist in FM and non-collinear AFMs, such as anomalous Hall effect[13-16], non-SOC spin current[17], magneto-optical responses[18], and C-paired spin-valley locking[19,20], etc. The joint advantages of FM-like transport and AFM-like ground state in AM open a new avenue for magnetic-based applications of message processing, data storage and even quantum computing, etc[21-24].

To have a comprehensive understanding of AM-related properties, realistic materials are desired for both materials computations and experimental measurements. With the concept in mind, many AM materials were theoretically proposed and some of them were experimentally verified[11,12,14,15,21-23,25-28]. In all these AMs, Rutile RuO$_2$ has a special position, since it was the first proposed AM with the largest spin split up to 1.4 eV[2]. So far, most of the interesting phenomena in AMs were carried by electron currents, which make realistic AM metals especially important and most related measurements were performed based on RuO$_2$. The collinear AFM magnetic structure in RuO$_2$ was identified based on the measurements of resonant X-ray scattering and neutron diffraction[29,30]. According to these reports, the two sublattices of Ru are connected by a four-fold screw rotation operation, belonging to the AM classification.

Very recently, a systematic study of the magnetic structure in RuO$_2$ single crystal was carried out via a local moment-sensitive measurement of muon spin rotation and relaxation (μSR) [31]. According to the new measurements, the local magnetic moment on each Ru site is in the order of 10$^{-4}$ μB, much smaller than the previous reports and AM requirements[30]. So, it can be considered as a nonmagnetic or paramagnetic ground state. It proposed that the previously reported and alleged magnetic reflections are likely related to multiple scattering artifacts rather than to a magnetic origin[32]. The conflicting evidence from different measurements motivated the research community to look back at all the AM-related transport properties in RuO$_2$. Therefore, it is crucial to have a clear and definitive understanding of its magnetic ground state. In this work we propose a strategy to identify the magnetic ground state based on the magneto-transport properties of quantum oscillation (QO). Recently, a systematic theoretical study of the QO for altermagnet were reported based on effective model analysis[33]. The big difference between the Fermi surfaces (FSs) in altermagnetic and nonmagnetic states of RuO$_2$ will lead to significant differences in the QO signals, which can be utilized as a powerful tool to understand the magnetic ground state[34].

## II. METHOD

The electronic structure calculations for RuO$_2$ were performed using the Vienna Ab initio Simulation Package (VASP)[35] with the Generalized Gradient Approximation (GGA) as formulated by Perdew, Burke, and Ernzerhof (PBE)[36]. The plane-wave basis set was employed with a kinetic energy cutoff of 500 eV to ensure the accuracy of the results. For the AM state calculations, a Hubbard U of 2 eV was applied, as previous DFT calculations and experimental measurements have shown that the magnetic moment and anomalous Hall conductivity (AHC) are best matched with experimental values when $U = 2$ eV[14,29,30,37,38]. After obtaining the converged electronic structure from VASP, we performed maximally localized Wannier function (MLWF) projection to construct the tight-binding Hamiltonians[39]. The Wannier functions were generated to represent the Ru d-orbitals and O p-orbitals, providing a detailed and accurate description of the electronic states near the Fermi level. Using the constructed tight-binding Hamiltonian, we calculated the extreme cross-sectional areas of the FSs. The Onsager relation[40], $F = (\frac{\hbar}{2\pi e})S_f$, was then employed to determine the QO frequencies ($F$). Oxygen defects are an inherent issue in transition metal oxides. Their presence introduces additional electrons, thereby elevating the Fermi level, which leads to slight changes in the quantum oscillation signals. Oxygen defects also act as scattering centers, reducing the mean free path of electrons, which can result in certain cyclotron orbits being undetectable in experiments. However, the overall structure of the Fermi surface and the fundamental characteristics of the magnetic ground state remain largely unaffected. So in this work, we considered the clean and low temperature approximations.

## III. RESULTS AND DISCUSSION

RuO$_2$ exhibits a rutile-type crystal structure, belonging to the space group *P*4$_2$/*mnm* (No. 136). In this structure, each Ru atom is surrounded by six O atoms, forming an octahedron with two-fold rotational symmetry. The crystal structure and magnetic structure are illustrated in Fig. 1(a) and (b), respectively. The potential magnetic moments are located on the Ru sites, originating from the 4d orbitals, with the easy axis along the [001] direction[30,41].

To investigate the differences between nonmagnetic and altermagnetic states of RuO$_2$, the FSs of both magnetic states are calculated, see Fig. 1(c-d). In the nonmagnetic state, there are three independent FS bubbles. The first one is close to the $k_z = \pm\pi/c$ plane (with label FS1), the second one is approximately spherical around the Γ point (with label FS2), and the third one is a small bubble located around the hinge of Brillouin zone (BZ) $k_x = k_y = \pm\pi/a$ (with label FS3). After taking the magnetic order into consideration with including on-site $U$=2 eV, the FSs change dramatically. As presented in Fig. 1(d), we plot the spin-up and spin-down channels connected by a S$_{4z}$ screw rotation symmetry in different colors. The magnetic RuO$_2$ FSs consist of three major parts: two intersecting semi-ellipsoids at the top and bottom of the BZ, FSs located at the four corners near $k_z = 0$, and a $d_{xy}$-orbital-like closed structure around the Γ point. The dramatic differences of FSs in the two possible magnetic states enable us to use the QO to identify the true ground states in experiments.

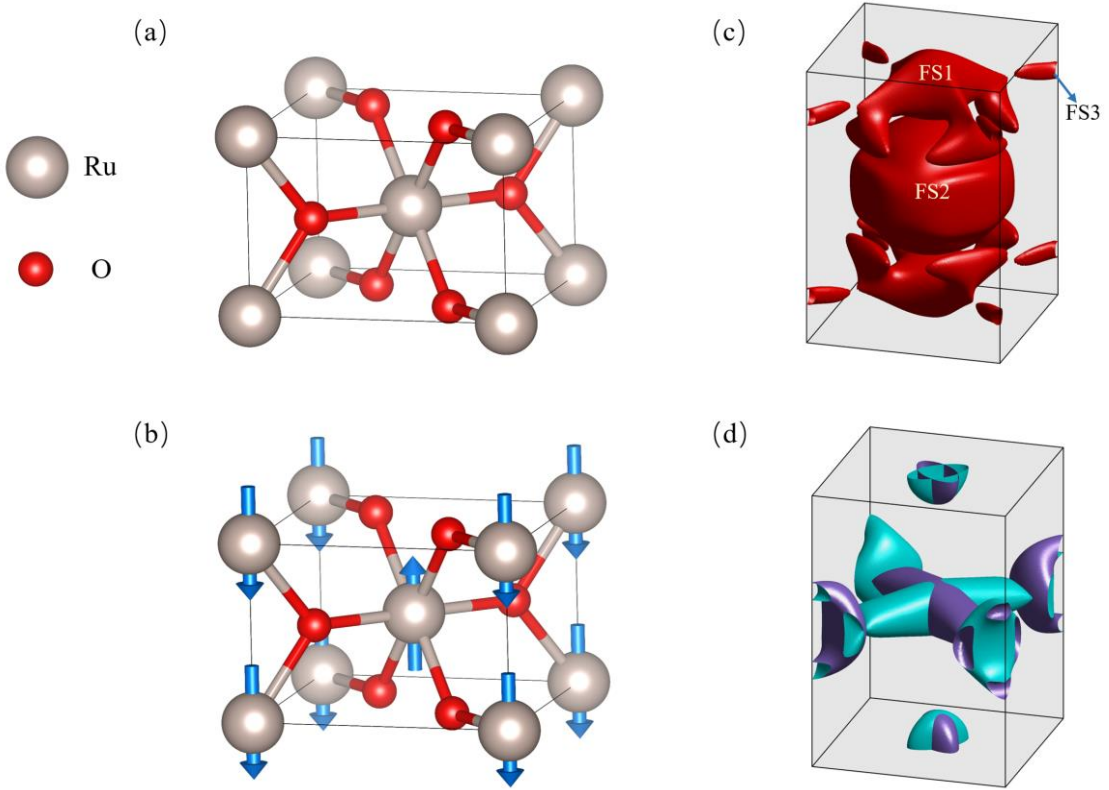

FIG.1. Crystal structure and Fermi surface (FS) of RuO$_2$. (a) Crystal structure of nonmagnetic RuO$_2$. (b) Crystal and magnetic structure of altermagnetic RuO$_2$. (c) FS of nonmagnetic RuO$_2$ with the consideration of spin-orbital coupling (SOC). (d) FS of altermagnetic RuO$_2$ without the consideration of SOC.

The difference between nonmagnetic and AM states can also be seen from electronic band structures, as presented in Fig. 2. From Fig. 2(b) one can see that the bands cutting Fermi level mainly happen around Γ points, on the line of Γ-X, Γ-M and Γ-Z. In addition, there is a small Fermi cutting on the line of A-M close to point A, fully consistent with the FS in Fig. 1(c). With the consideration of spin-orbital coupling (SOC), some degeneracy on Γ-Z, R-Z, and R-A are broken but just with small spin splittings, due to weak SOC induced the SU(2) symmetry breaking. After taking magnetic order into consideration, the electronic band structure changes completely, and a large spin split around 1.4eV can be found in the M-Γ direction near the Fermi level, see Fig. 2(c), fully consistent with previous reports[2]. Similar to the case without adding on-site U, some new band splittings are generated by the SOC effect, which leads to a visible dispersion of the original flat band in the spin-down channel along M-Γ. Owing to the changes of energy dispersions after considering the magnetic moment, the Fermi cutting also changes a lot. In comparison to the case without magnetism, the distances from Fermi cutting to Γ become very different on the Γ-X, Γ-M, and Γ-Z, leading to a $d_{xy}$-orbital-like flat FS around Γ points. Besides, one new small FS is formed around the M point for each spin channel.

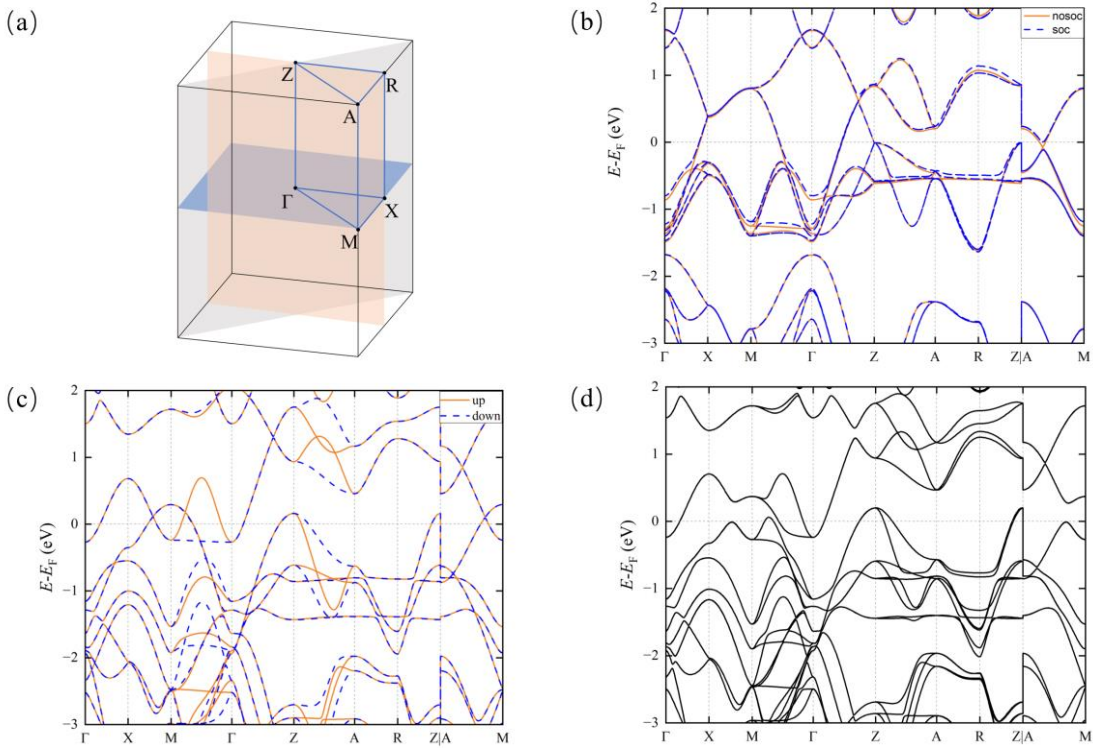

FIG.2. The Brillouin zone (BZ) and band structure of $RuO_2$. (a) The three-dimensional BZ of $RuO_2$, with the blue plane representing the [001] plane, the orange plane representing the [100] plane, and the gray plane representing the [110] plane. (b) Band structure of nonmagnetic $RuO_2$ without considering SOC and with considering SOC, respectively. (c) Band Structure of altermagnetic $RuO_2$ without considering SOC (orange solid line for spin up, the blue dash for spin down). (d) Band Structure of altermagnetic $RuO_2$ with considering SOC.

To simulate the QO in magneto-transport signals, we adopted an angular rotation strategy to provide as many characteristic curves of QO frequencies as possible, thereby offering more robust evidence for diagnosing the presence of magnetism through QO measurements. The rotation angles span from the [001] direction to the [100] direction by 90°, then from the [100] direction to the [110] direction by 45°, and finally from the [110] direction back to the [001] direction by 90°. This extensive angular range allows us to collect nearly comprehensive QO frequency data, providing a thorough basis for our analysis. Though the ground state can be understood without considering SOC, SOC always exists in practical materials, and its effect is not small in a 4d-orbital dominated FS here. So, in all the QO calculations, we included the SOC effect.

Fig. 3(a) shows the angle-dependent QO frequencies of nonmagnetic $RuO_2$, derived from the extreme cross-sectional areas of three independent closed FSs. The evolution of the angle-dependent QO contributed from FS1 can be effectively described by 11 typical extreme cross-sections along with the rotation of the applied magnetic field, as shown in Fig. 3(b).

Initially, as the magnetic field rotates from the [001] direction to the [100] direction in the range of 0° to ~24°, two maximums appear at the front legs of FS1, corresponding to the FS1-cut1 curve. After 24°, no extreme value exists anymore. Though FS1 follows the four-fold rotation symmetry, the contribution from different legs of this FS is different when we rotate the magnetic field from [001] to [100]. The contribution of the other two legs can survive in a relatively large range of 0° to ~60°. So, we can see that the four legs of FS1 correspond to frequencies close to 900T with magnetic field rotating from [001] to [100], see the left lower corner in Fig. 3(a). Following FS1-cut1 at the angle of ~24°, further rotating the magnetic field, a new cross section appears, see the FS1-cut3 in Fig. 3(b). This extreme cross section can exist in a large range of ~25° to 90° and can even exist in a small range of 25° to ~6° in the other rotation direction from [100] to [110]. The extreme cross section of FS1-cut3 can contribute to a large Frequency window in the range of ~2400T to ~4600T, see the label on the left part of Fig. 3(a).

When we rotate the magnetic field to ~75°, a new extreme cross section appears, see FS1-cut4 in Fig. 3(b). Similar to the case in FS1-cut3, the extreme cross section from FS1-cut4 can also exist in the two rotation directions from [001] to [100] and from [100] to [110], with the frequencies in a range of ~2100T to ~2700T, see the corresponding label in Fig. 3(a). Around the [110] direction, a large extreme area involving two nearby legs and the head of FS1 appears, see FS1-cut5 in Fig. 3(b). It can survive in the range up to ~15° and ~70° with rotating magnetic field away from [110] to [100] and [001], respectively, with frequencies in the range of ~3600T to ~4700T. The front right leg of FS1 can generate another effective cutting as FS1-cut6 with frequencies near ~1000T when the magnetic field is rotating between [100] and [110] directions.

A smaller section from FS1-cut7 represents the connection between the front right leg and the head, see Fig. 3(b). As rotation passes the [110] direction and continues towards the [001] direction, FS1-cut7 evolves into two parts, with FS1-cut8 from the rear left leg and head connection and FS1-cut9 from the front right leg and head connection. These two minimums exhibit entirely different trends with rotation, where FS1-cut9 increases sharply from ~400T to ~1800T and FS1-cut8 remains almost constant near ~300T.

Similarly, as rotation passes the [110] direction and continues towards the [001] direction, FS1-cut6 evolves into two parts, with FS1-cut10 from the rear left leg and FS1-cut11 from the front right leg. These two maximums also exhibit different trends with rotation, where FS1-cut10 increases sharply from ~1400T to ~3000T and FS1-cut11 decreases from ~1400T to ~900T.

In comparison to FS1, the shape of FS2 is much simpler in a form that's almost spherical. So, the corresponding evolution of frequencies along with rotation is continuous in all the checked directions and confined in the range of ~8500T to ~10300T, as presented in the top part of Fig. 3(a). In addition, there is also a very small FS3 located around the corner of BZ with a "cross-like" shape, see Fig.3 (d). Frequency evolution of FS3 is also continuous in all the checked directions, as presented at the bottom of Fig. 3(a). The extreme cross-sectional area corresponding to FS3-cut ranges from ~100T to ~800T, with the continuous evolution of frequency across the whole angle.

Fig. 4(a) shows the angle-dependent QO frequencies of AM phase $RuO_2$, with the magnetic field rotated through the same angles as in the nonmagnetic case. There are five independent FS bubbles in total, as presented in Fig. 4(b). The extreme cross-sections of the magnetic $RuO_2$ FSs (FS1-FS4) are relatively simpler and more straightforward due to their convex FS, exhibiting continuous evolution throughout the rotation. FS1-cut has its smallest extreme cross-section along the [001] direction, gradually reaching its maximum as it rotates to [100] direction, with continuous frequencies in the range of ~700T to ~1400T. FS2-cut has a nearly square minimum cross-section in the [001] direction and reaches its maximum when rotated to the [100] direction, with frequencies in the range of ~1000T to ~1400T. FS3-cut reaches a minimum at the [100] direction and a maximum at the [110] direction with frequencies in the range of ~1500T to ~2000T. FS4-cut has an "x"-shaped cross-section along the [001] direction, corresponding to the maximum QO frequency ~6000T. When rotated to the [100] direction, FS4-cut reaches a frequency minimum at ~1700T.

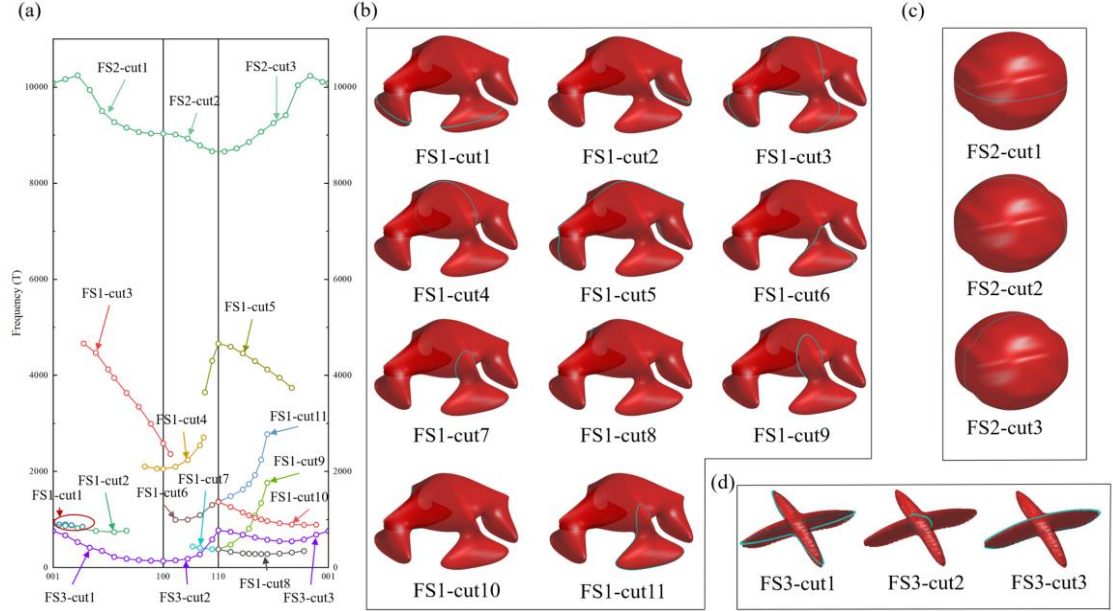

FIG.3. The angle-dependent quantum oscillation (QO) of $RuO_2$ in nonmagnetic phase and the cross-sectional extreme value details. (a) Angle-dependent QO frequency with applied magnetic field changing from [001] to [100], [100] to [110], and finally from [110] back to [001]. (b-d) Typical extreme cross-sectional areas in FS1, FS2, and FS3.

FS5 is relatively complicated and has three distinct extreme cross-sections. As the magnetic field rotates from the [001] direction to the [100] direction in the range of ~60° to ~75°, two maximums appear in FS5, corresponding to the curve of FS5-cut1 in Fig.4(b). After ~85°, these two maximums merge to form a new maximum, see FS5-cut2 in Fig. 4(b). This cross section survives in the whole range of [100] to [110] and continues from 0° to ~75° from [110] to [001], with frequencies from ~2200T to ~2600T. After that, the one big cross section becomes four small maximums, see FS5-cut3 in Fig. 4(b), with frequencies near ~500T, see FS5-cut3 in Fig. 4(b).

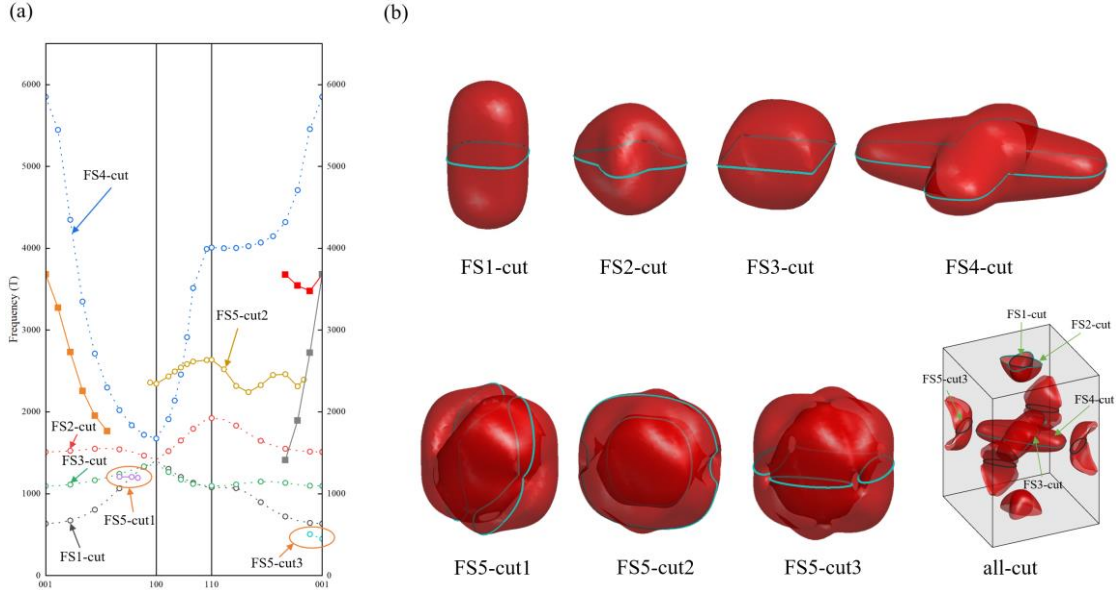

FIG.4. The angle-dependent QO frequencies of altermagnetic $RuO_2$ and the cross-sectional extreme value details. (a) Angle-dependent QO frequency data. The direction of the applied magnetic field changes from [001] to [100], [100] to [110], and finally from [110] back to [001]. (b) Typical extreme cross-sectional areas in FS1-FS5.

We considered the presence of SOC in real materials and further mapped the spin texture for each cross-section of the altermagnetic $RuO_2$, as shown in Fig. 5. It can be observed in Fig. 5(a) and Fig. 5(e) that there are sharp spin transitions at certain points, which are generally thought to prevent electrons from smoothly passing through, making it difficult to observe these QO frequencies in experiments. In Fig. 4(a), the QO frequencies corresponding to sharp spin transitions are represented by dotted lines, while those with continuous spin variations are indicated by solid lines.

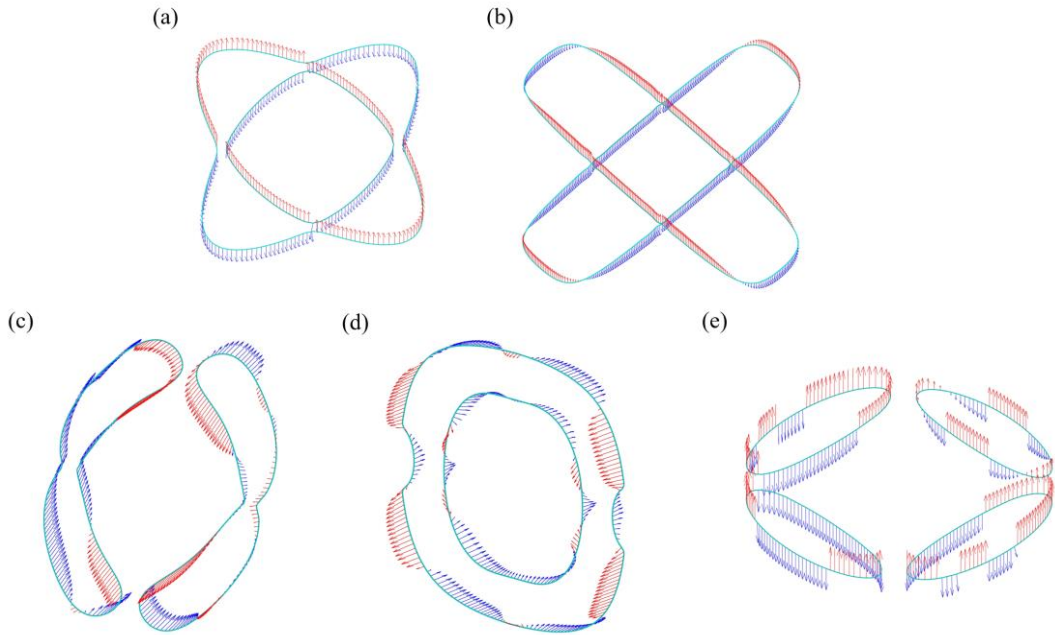

FIG.5. The spin texture of various cross-sections corresponding to the altermagnetic $RuO_2$. (a) Spin texture of FS1-cut and FS2-cut, (b) Spin texture of FS3-cut and FS4-cut, (c-e) Spin texture of FS5 (red represents spin along the positive direction of the cross-section, blue arrows represent spin along the negative direction of the cross-section).

Interestingly, in the [001] direction, FS3 and FS4 exhibit a topological structure of the Fermi surface, where the Fermi surfaces of these two bands touch within the red box. This occurs due to the protection by mirror symmetry. As shown in Fig. 6(b), when an electron initially follows the green trajectory, due to a sharp change in its spin, this trajectory may not be detectable in experiments. Instead, as the electron approaches the nodal point, magnetic breakdown occurs, causing the electron to enter a different orbit, consistent with the principle of continuous spin evolution. This alters the original cyclotron orbit and results in different QO frequencies.

The change of area is illustrated in Fig. 6(c). The top two cross sections show the original electron orbits, while the bottom two are the new orbits after the switch of trajectory. The cross-sectional areas of the original orbits are 0.5581 Å$^{-2}$ and 0.1044 Å$^{-2}$, corresponding to two new areas of 0.3316 Å$^{-2}$. This leads to the appearance of new QO frequencies, which are represented by the squares on the left and right parts of Fig. 4(a).

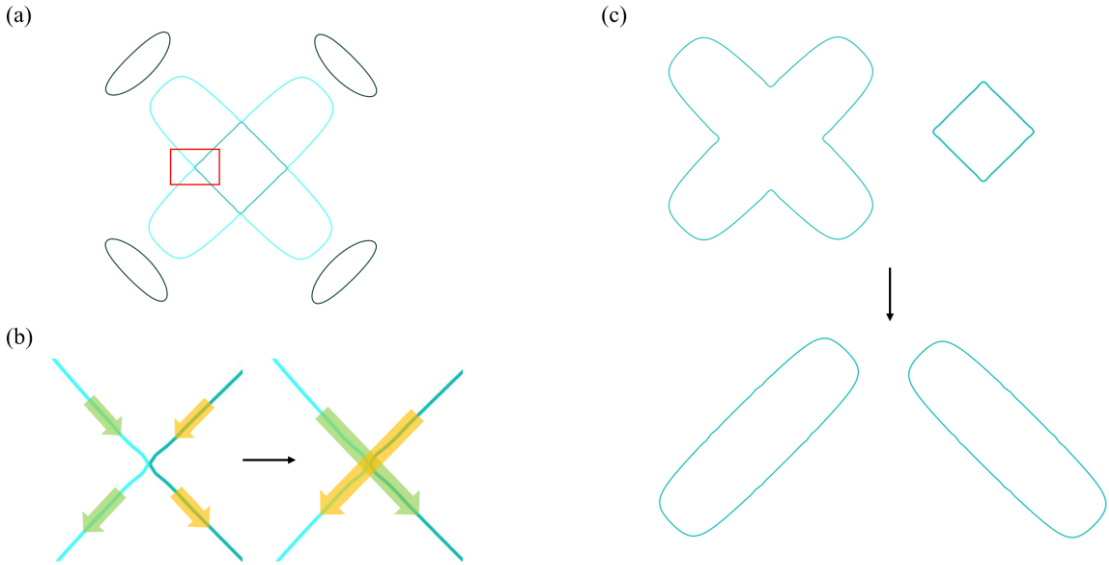

FIG.6. Topological structure of the cross section when magnetic field towards [001] in AM phase. (a) The FS cross section in the BZ when magnetic field towards [001]. The topological structure appears in the red box. (b) The topological structure of the FS changes the trajectory of electron. (c) Different electron cyclotron orbits appear due to the topological structure of the FS.

Due to the topological structure being symmetry-protected, rotating the magnetic field direction causes the separation of the FS contact points. However, because of magnetic breakdown[42,43], when electrons move along cyclotron orbits, they can "tunnel" through small gaps, resulting in new frequencies from complex trajectories[44]. The magnetic breakdown probability occurring when electrons pass through these nodes can be expressed as:

$$P \approx e^{-B_0/B}, \quad \text{where} \quad B_0 = \frac{\hbar k_g^2}{e}$$

where $B_0$ is the characteristic breakdown field, and $B$ is the applied external magnetic field. The value of $B_0$ is determined by the breakdown gap $k_g$ between the two trajectories in k-space[45]. Here, $\hbar$ is the reduced Planck constant, and $e$ is the electron charge. Under typical experimental conditions, $P$ ranges between 0 and 1. We performed fine angular adjustments near the [001] direction and found that the probability of magnetic breakdown gradually decreases with small angular changes, reaching approximately 35% at around 10°. The angle-dependent magnetic breakdown probability is shown in Table I. This phenomenon indicates that the new frequency does not immediately vanish with slight changes in the magnetic field direction even if the symmetry is broken.

TABLE I. This table shows the variation of magnetic breakdown probability with the angle θ in two directions away from [001], under an external magnetic field $B$ of 10T.

| [001]-[100] (θ) | 1° | 2° | 3° | 4° | 6° | 8° | 10° |
|---|---|---|---|---|---|---|---|
| $P(B=10T)$ | 94.0% | 91.2% | 87.9% | 83.0% | 70.3% | 55.3% | 35.0% |
| [001]-[110] (θ) | 1° | 2° | 3° | 4° | 6° | 8° | 10° |
| $P(B=10T)$ | 94.7% | 93.8% | 92.2% | 90.7% | 86.3% | 76.9% | 35.9% |

## IV. SUMMARY

In summary, we have presented a detailed analysis for the QO in the two most promising ground states of nonmagnetic and AM phases. Our calculations show dramatic differences of their FSs and corresponding QO frequency diagrams, which can be easily identified by magneto-transport measurements. We believe that it is sufficient to determine the presence or absence of magnetism in $RuO_2$ by this method. Moreover, we also found a signature of topological structure in the FS cross-section in the [001] direction of magnetic $RuO_2$. This symmetry-protected nodal line band structure can lead to some new electron trajectories and QO frequencies, which can serve as another important feature for the identification.

## ACKNOWLEDGMENTS

This work was supported by the National Key R&D Program of China 2021YFB3501503, National Natural Science Foundation of China (Grants No.52271016, No.52188101), and Liaoning Province (Grant No. XLYC2203080). Part of the numerical calculations in this study were carried out on the ORISE Supercomputer (No. DFZX202319).